\errorstopmode
\input amssym.def
\input amssym.tex


\magnification=\magstephalf
\hsize=14.0 true cm
\vsize=19 true cm
\hoffset=1.0 true cm
\voffset=2.0 true cm

\abovedisplayskip=12pt plus 3pt minus 3pt
\belowdisplayskip=12pt plus 3pt minus 3pt
\parindent=1.0em


\font\sixrm=cmr6
\font\eightrm=cmr8
\font\ninerm=cmr9

\font\sixi=cmmi6
\font\eighti=cmmi8
\font\ninei=cmmi9

\font\sixsy=cmsy6
\font\eightsy=cmsy8
\font\ninesy=cmsy9

\font\sixbf=cmbx6
\font\eightbf=cmbx8
\font\ninebf=cmbx9

\font\eightit=cmti8
\font\nineit=cmti9

\font\eightsl=cmsl8
\font\ninesl=cmsl9

\font\sixss=cmss8 at 8 true pt
\font\sevenss=cmss9 at 9 true pt
\font\eightss=cmss8
\font\niness=cmss9
\font\tenss=cmss10

\font\sixmib=cmmib6
\font\sevenmib=cmmib7
\font\eightmib=cmmib8
\font\ninemib=cmmib9
\font\tenmib=cmmib10

 at 12 true pt
 at 12 true pt
\font\bigrm=cmr10 at 12 true pt
 at 12 true pt
 at 12 true pt

 at 16 true pt
 at 16 true pt
\font\Bigrm=cmr12 at 16 true pt
 at 16 true pt
 at 16 true pt

\catcode`@=11
\newfam\ssfam
\newfam\mibfam

\def\tenpoint{\def\rm{\fam0\tenrm}%
    \textfont0=\tenrm \scriptfont0=\sevenrm \scriptscriptfont0=\fiverm
    \textfont1=\teni  \scriptfont1=\seveni  \scriptscriptfont1=\fivei
    \textfont2=\tensy \scriptfont2=\sevensy \scriptscriptfont2=\fivesy
    \textfont3=\tenex \scriptfont3=\tenex   \scriptscriptfont3=\tenex
    \textfont\itfam=\tenit                  \def\it{\fam\itfam\tenit}%
    \textfont\slfam=\tensl                  \def\sl{\fam\slfam\tensl}%
    \textfont\bffam=\tenbf \scriptfont\bffam=\sevenbf
                           \scriptscriptfont\bffam=\fivebf
                           \def\bf{\fam\bffam\tenbf}%
    \textfont\ssfam=\tenss \scriptfont\ssfam=\sevenss
                           \scriptscriptfont\ssfam=\sevenss
                           \def\ss{\fam\ssfam\tenss}%
    \textfont\mibfam=\tenmib \scriptfont\mibfam=\sevenmib
                             \scriptscriptfont\mibfam=\sevenmib
                             \def\mib{\fam\mibfam\tenmib}%
    \normalbaselineskip=13pt
    \setbox\strutbox=\hbox{\vrule height8.5pt depth3.5pt width0pt}%
    \let\big=\tenbig
    \normalbaselines\rm}

\def\ninepoint{\def\rm{\fam0\ninerm}%
    \textfont0=\ninerm      \scriptfont0=\sixrm
                            \scriptscriptfont0=\fiverm
    \textfont1=\ninei       \scriptfont1=\sixi
                            \scriptscriptfont1=\fivei
    \textfont2=\ninesy      \scriptfont2=\sixsy
                            \scriptscriptfont2=\fivesy
    \textfont3=\tenex       \scriptfont3=\tenex
                            \scriptscriptfont3=\tenex
    \textfont\itfam=\nineit \def\it{\fam\itfam\nineit}%
    \textfont\slfam=\ninesl \def\sl{\fam\slfam\ninesl}%
    \textfont\bffam=\ninebf \scriptfont\bffam=\sixbf
                            \scriptscriptfont\bffam=\fivebf
                            \def\bf{\fam\bffam\ninebf}%
    \textfont\ssfam=\niness \scriptfont\ssfam=\sixss
                            \scriptscriptfont\ssfam=\sixss
                            \def\ss{\fam\ssfam\niness}%
    \textfont\mibfam=\ninemib \scriptfont\mibfam=\sixmib
                            \scriptscriptfont\mibfam=\sixmib
                            \def\mib{\fam\mibfam\ninemib}%
    \normalbaselineskip=12pt
    \setbox\strutbox=\hbox{\vrule height8.0pt depth3.0pt width0pt}%
    \let\big=\ninebig
    \normalbaselines\rm}

\def\eightpoint{\def\rm{\fam0\eightrm}%
    \textfont0=\eightrm      \scriptfont0=\sixrm
                             \scriptscriptfont0=\fiverm
    \textfont1=\eighti       \scriptfont1=\sixi
                             \scriptscriptfont1=\fivei
    \textfont2=\eightsy      \scriptfont2=\sixsy
                             \scriptscriptfont2=\fivesy
    \textfont3=\tenex        \scriptfont3=\tenex
                             \scriptscriptfont3=\tenex
    \textfont\itfam=\eightit \def\it{\fam\itfam\eightit}%
    \textfont\slfam=\eightsl \def\sl{\fam\slfam\eightsl}%
    \textfont\bffam=\eightbf \scriptfont\bffam=\sixbf
                             \scriptscriptfont\bffam=\fivebf
                             \def\bf{\fam\bffam\eightbf}%
    \textfont\ssfam=\eightss \scriptfont\ssfam=\sixss
                             \scriptscriptfont\ssfam=\sixss
                             \def\ss{\fam\ssfam\eightss}%
    \textfont\mibfam=\eightmib \scriptfont\mibfam=\sixmib
                             \scriptscriptfont\mibfam=\sixmib
                             \def\mib{\fam\mibfam\eightmib}%
    \normalbaselineskip=10pt
    \setbox\strutbox=\hbox{\vrule height7.0pt depth2.0pt width0pt}%
    \let\big=\eightbig
    \normalbaselines\rm}

\def\tenbig#1{{\hbox{$\left#1\vbox to8.5pt{}\right.\n@space$}}}
\def\ninebig#1{{\hbox{$\textfont0=\tenrm\textfont2=\tensy
                       \left#1\vbox to7.25pt{}\right.\n@space$}}}
\def\eightbig#1{{\hbox{$\textfont0=\ninerm\textfont2=\ninesy
                       \left#1\vbox to6.5pt{}\right.\n@space$}}}

\font\sectionfont=cmbx10
\font\subsectionfont=cmti10

\def\figurecaptionfont{\ninepoint}
\def\tablecaptionfont{\ninepoint}


\newcount\equationno
\newcount\bibitemno
\newcount\figureno
\newcount\tableno

\equationno=0
\bibitemno=0
\figureno=0
\tableno=0


\footline={\ifnum\pageno=0{\hfil}\else
{\hss\rm\the\pageno\hss}\fi}


\def\section #1. #2 \par
{\vskip0pt plus .10\vsize\penalty-100 \vskip0pt plus-.10\vsize
\vskip 1.6 true cm plus 0.2 true cm minus 0.2 true cm
\global\def\equationlabel{#1}
\global\equationno=0
\leftline{\sectionfont #1. #2}\par
\immediate\write\terminal{Section #1. #2}
\vskip 0.7 true cm plus 0.1 true cm minus 0.1 true cm
\noindent}


\def\subsection #1 \par
{\vskip0pt plus 0.8 true cm\penalty-50 \vskip0pt plus-0.8 true cm
\vskip2.5ex plus 0.1ex minus 0.1ex
\leftline{\subsectionfont #1}\par
\immediate\write\terminal{Subsection #1}
\vskip1.0ex plus 0.1ex minus 0.1ex
\noindent}


\def\appendix #1. #2 \par
{\vskip0pt plus .10\vsize\penalty-100 \vskip0pt plus-.10\vsize
\vskip 1.6 true cm plus 0.2 true cm minus 0.2 true cm
\global\def\equationlabel{\hbox{\rm#1}}
\global\equationno=0
\leftline{\sectionfont Appendix #1. #2}\par
\immediate\write\terminal{Appendix #1. #2}
\vskip 0.7 true cm plus 0.1 true cm minus 0.1 true cm
\noindent}



\def\equation#1{$$\displaylines{\qquad #1}$$}
\def\enum{\global\advance\equationno by 1
\hfill\llap{{\rm(\equationlabel.\the\equationno)}}}
\def\noenum{\hfill}

\def\nexteq#1{\cr\noalign{\vskip#1}\qquad}


\def\ifundefined#1{\expandafter\ifx\csname#1\endcsname\relax}

\def\ref#1{\ifundefined{#1}?\immediate\write\terminal{unknown reference
on page \the\pageno}\else\csname#1\endcsname\fi}

\newwrite\terminal
\newwrite\bibitemlist

\def\bibitem#1#2\par{\global\advance\bibitemno by 1
\immediate\write\bibitemlist{\string\def
\expandafter\string\csname#1\endcsname
{\the\bibitemno}}
\item{[\the\bibitemno]}#2\par}

\def\beginbibliography{
\vskip0pt plus .15\vsize\penalty-100 \vskip0pt plus-.15\vsize
\vskip 1.2 true cm plus 0.2 true cm minus 0.2 true cm
\leftline{\sectionfont References}\par
\immediate\write\terminal{References}
\immediate\openout\bibitemlist=biblist
\frenchspacing\parindent=1.8em
\vskip 0.5 true cm plus 0.1 true cm minus 0.1 true cm}

\def\endbibliography{
\immediate\closeout\bibitemlist
\nonfrenchspacing\parindent=1.0em}


\def\figurecaption#1{\global\advance\figureno by 1
\narrower\figurecaptionfont
Fig.~\the\figureno. #1}

\def\tablecaption#1{\global\advance\tableno by 1
\vbox to 0.25 true cm { }
\centerline{\tablecaptionfont%
Table~\the\tableno. #1}
\vskip-0.4 true cm}

\def\thintablerule{\hrule height0.4pt}

\tenpoint

\immediate\openin\bibitemlist=biblist
\ifeof\bibitemlist\immediate\closein\bibitemlist
\else\immediate\closein\bibitemlist
\input biblist \fi


\def\thismonth{\ifcase\month\or
January\or February\or March\or April\or May\or June\or
July\or August\or September\or October\or November\or December\fi}

\input epsf
\epsfclipon



\def\rmd{{\rm d}}
\def\rmD{{\rm D}}
\def\rme{{\rm e}}
\def\rmO{{\rm O}}


\def\rz{{\Bbb R}}
\def\gz{{\Bbb Z}}


\def\proof{\noindent{\sl Proof:}\kern0.6em}

\def\frac#1#2{\hbox{$#1\over#2$}}
\def\dual{\mathstrut^*\kern-0.1em}

\def\lvec#1{\setbox0=\hbox{$#1$}
    \setbox1=\hbox{$\scriptstyle\leftarrow$}
    #1\kern-\wd0\smash{
    \raise\ht0\hbox{$\raise1pt\hbox{$\scriptstyle\leftarrow$}$}}
    \kern-\wd1\kern\wd0}
\def\rvec#1{\setbox0=\hbox{$#1$}
    \setbox1=\hbox{$\scriptstyle\rightarrow$}
    #1\kern-\wd0\smash{
    \raise\ht0\hbox{$\raise1pt\hbox{$\scriptstyle\rightarrow$}$}}
    \kern-\wd1\kern\wd0}
\def\slash#1{\setbox0=\hbox{$#1$}\setbox1=\hbox{$\kern1pt/$}
    #1\kern-\wd0\kern1pt/\kern-\wd1\kern\wd0}


\def\nabstar#1{{\nabla\kern0.5pt\smash{\raise 4.5pt\hbox{$\ast$}}
               \kern-5.5pt_{#1}}}

\def\drvstar#1{{\partial\kern0.5pt\smash{\raise 4.5pt\hbox{$\ast$}}
               \kern-6.0pt_{#1}}}

\def\ldrvstar#1{{\lvec{\,\partial}\kern-0.5pt\smash{\raise 4.5pt\hbox{$\ast$}}
               \kern-5.0pt_{#1}}}


\def\MSbar{\overline{\rm MS\kern-0.5pt}\kern0.5pt}


\def\euler{\gamma_{\rm E}}



\def\diracstar#1#2{
    \setbox0=\hbox{$\gamma$}\setbox1=\hbox{$\gamma_{#1}$}
    \gamma_{#1}\kern-\wd1\kern\wd0
    \smash{\raise4.5pt\hbox{$\scriptstyle#2$}}}


\def\tr{{\rm tr}}
\def\Tr{{\rm Tr}}
\def\Ad{{\rm Ad}\,}
\def\Group{{\rm SU}(3)}
\def\Lie{\frak{su}(3)}


\def\trans{{\cal F}}
\def\transs{\trans_{*}}
\def\transt{\trans_{t}}
\def\transts{\trans_{t,*}}
\def\transne{\trans_{n\eps}}
\def\transnes{\trans_{n\eps,*}}
\def\euler#1{{\cal E}_{#1}}
\def\du{\partial}


\def\Sw{S_{\rm w}}
\def\Sflow{\widetilde{S}}
\def\Wl{{\cal W}}
\def\Proj{{\cal P}}
\def\Lop{\frak{L}_t}
\def\Lap{\frak{L}_0}
\def\Lopz{\frak{L}_{t_0}}


\def\mom{\pi}
\def\eps{\epsilon}


\def\Cinfty{C^{\infty}}
\def\Gop{{\frak G}_t}
\def\Gopz{{\frak G}_{t_0}}
\def\Pop{{\frak P}_t}
\def\Popz{{\frak P}_{t_0}}
\def\Lopp{{\frak L}'}

%
\rightline{CERN-PH-TH/2009-118}

\vskip1.5cm 
\centerline{\Bigrm
Trivializing maps, the Wilson flow and}
\vskip0.3cm
\centerline{\Bigrm
the HMC algorithm}
\vskip 0.6 true cm
\centerline{\bigrm Martin L\"uscher}
\vskip1.5ex
\centerline{{\it CERN, Physics Department, 1211 Geneva 23, Switzerland}}
\vskip 0.8 true cm
\thintablerule
\vskip 2.0ex
\ninepoint
\leftline{\bf Abstract}
\vskip 1.0ex\noindent
In lattice gauge theory, there exist field transformations 
that map the theory to the trivial one, where the
basic field variables
are completely decoupled from one another.
Such maps can be constructed systematically
by integrating certain flow equations 
in field space. 
The construction is worked out
in some detail and it is proposed to
combine the Wilson flow 
(which generates approximately trivializing maps
for the Wilson gauge action) 
with the HMC simulation algorithm
in order to improve the efficiency of 
lattice QCD simulations.
\vskip 2.0ex
\thintablerule

\tenpoint

\vskip-0.3cm

\section 1. Introduction

The Nicolai map  
transforms interacting supersymmetric theories
to non-interacting ones [\ref{NicolaiMap}].
Supersymmetry is considered to be essential for 
the existence of these field transformations in view
of the fact that their Jacobian is exactly 
cancelled by the fermion partition function.

In lattice gauge theory,
a natural question to ask is
whether there are field transformations that map
the theory to its strong-coupling limit.
In particular, if there are no matter fields, 
one is looking for substitutions
\equation{
  U=\trans(V)
  \enum
} 
of the gauge field $U$ in the functional integral 
whose Jacobian cancels the gauge-field action.
Similarly to the Nicolai map, this kind of
transformation maps the theory to a solvable one, but
supersymmetry is not required and the Jacobian
plays a different r\^ole.

On a finite lattice, and if the gauge group is compact
and connected, the existence of such 
trivializing maps is guaranteed by 
a general theorem on volume forms on compact manifolds
(see ref.~[\ref{MoserZehnder}], Theorem 1.26, for example).
One may be inclined to assume that these transformations 
are too complicated to be of any use.
However, as explained later in this paper,
it is possible to build up trivializing maps
by integrating flows in field space, whose generators satisfy
certain partial differential equations.
The latter are quite tractable and
can, to some extent, be solved analytically in the pure
gauge theory. An application of trivializing
maps, which can then be envisaged, is the
acceleration of lattice QCD simulations.

\topinsert
\vbox{
\vskip0.0cm
\centerline{\epsfxsize=6.0cm\epsfbox{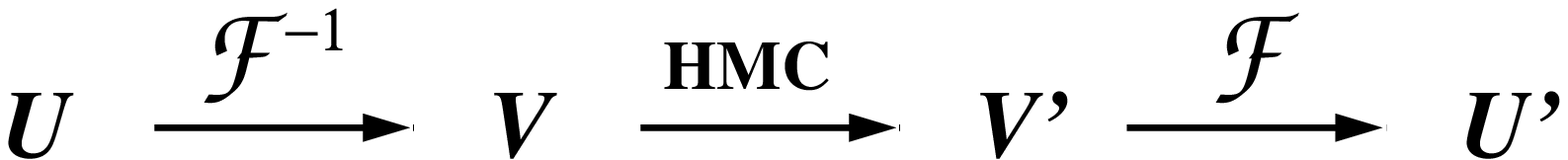}}
\vskip0.5cm
\figurecaption{%
The proposed
simulation algorithm for lattice QCD
updates the gauge field $U$ in three
steps, following the arrows in this diagram, where
the Hamilton function used in the HMC step
has the standard kinetic term and includes the Jacobian
of the field transformation $\cal F$
(cf.~subsect.~2.4).
}
}
\endinsert

The fact that the efficiency of the 
available simulation algorithms is unpredictable
has always been a weakness of numerical lattice QCD.
Already a while ago, empirical studies 
of the $\Group$ gauge theory by
Del Debbio, Panagopoulos and Vicari [\ref{DelDebbioTauQ}]
showed that the
autocorrelation times of observables related to the topological
charge of the gauge field tend to be large and 
appear to grow exponentially with the inverse of the lattice spacing.
Moreover, 
Schaefer, Sommer and Virotta [\ref{StefanConference}] 
recently found that the situation is, in this respect, essentially
unchanged when the sea quarks are included in the simulations.

The rapid slowdown of the simulations at small lattice spacings
may conceivably be overcome
by combining approximately trivializing maps
with the HMC simulation algorithm [\ref{HMC}]
(see fig.~1).
Since the transformation moves the theory
closer to the strong-coupling limit, where
the HMC algorithm is known to be highly efficient,
the autocorrelation times are, in general, expected to be
reduced in this way. Evidently, for the combined algorithm
to work out in practice, approximately trivializing maps
must be found which are fairly simple and programmable.
One of the goals of the present paper is thus to provide a 
solution to this problem.

\vfill\eject

\section 2. Field transformations

Most concepts developed in this paper
are expected to be widely applicable, but as explained above, 
the case of immediate interest is lattice QCD. 
In the following, the gauge group is therefore taken to 
be $\Group$.
Since the quarks will play a spectator r\^ole,
they will be added to the theory only in sect.~6, where the proposed 
simulation algorithm for lattice QCD is discussed.

\subsection 2.1 Field space

The lattice theory is set up on a finite 
hypercubic lattice $\Lambda$ with periodic boundary conditions.
For notational convenience, the lattice spacing is set to unity.
As usual, the gauge field variables $U(x,\mu)\in\Group$
are assumed to
reside on the links $(x,\mu)$ of the lattice 
(where $x\in\Lambda$ and $\mu=0,\ldots,3$).
The expectation value of any observable
${\cal O}(U)$ is then given by the functional integral
\equation{
  \langle{\cal O}\rangle=
  {1\over{\cal Z}}\int\rmD[U]\,{\cal O}(U)\,\rme^{-S(U)}
  \enum
}
over the space of gauge fields. In this expression,
$S(U)$ denotes the gauge-field action, 
${\cal Z}$ the partition function and
$\rmD[U]$ the product of the normalized 
$\Group$-invariant integration measures of the link variables
$U(x,\mu)$.

From a purely mathematical point of view, the space of 
lattice gauge fields
is a power of $\Group$ and therefore a compact connected manifold.
Field transformations 
are invertible maps of this manifold onto itself.
Such transformations will always be required to be
differentiable 
in both directions and orientation-preserving
(here and below, ``differentiable'' means
``infinitely often continuously differentiable'').

\subsection 2.2 Right-invariant differential operators

Since the link variables $U(x,\mu)$ take values in a Lie group,
it is natural to express differentiations with respect
to them through a basis
$\du^a_{x,\mu}$, $a=1,\ldots,8$, of
differential operators that are invariant under the right-action
of the group. 
The action of these operators 
on a differentiable function $f(U)$ of the gauge field
is given by
\equation{
  \du_{x,\mu}^af(U)=\left.{\rmd\over\rmd t}
  f(U_t)\right|_{t=0},\hskip1.7em
  U_t(y,\nu)=\cases{\rme^{tT^a}U(x,\mu)& if $(y,\nu)=(x,\mu)$,\cr
                              \noalign{\vskip1.3ex}
                              U(y,\nu)           & otherwise,\cr}
  \enum
}
where $T^a$ are the $\Group$ generators (see appendix A).
In particular,
$\du^a_{x,\mu}$ transforms according to the adjoint
representation under the left-action of $\Group$.

The operators $\du^a_{x,\mu}$ go along with
a basis of 1-forms on the field manifold,
\equation{
  \theta^a_{x,\mu}=-2\,\tr\{\rmd U(x,\mu)U(x,\mu)^{-1}T^a\},
  \enum
}
such that
\equation{
  \rmd f(U)=\sum_{x,\mu}\theta^a_{x,\mu}\du^a_{x,\mu}f(U)
  \enum
}
for all functions $f(U)$.

\subsection 2.3 Jacobian matrix

For any given gauge field $V(y,\nu)$,
the transformation (1.1) produces another field
$U(x,\mu)=[\trans(V)](x,\mu)$.
When considering such transformations, one
needs to distinguish differentiations with respect to $V$
from those with respect $U$. The associated 1-forms must
also be distinguished. In the following, all symbols
carrying a hat represent quantities and operations
referring to $V$.

The Jacobian matrix
\equation{
  [\transs(V)](x,\mu;y,\nu)^{ab}=
  -2\,\tr\{\hat{\du}^b_{y,\nu}U(x,\mu)U(x,\mu)^{-1}T^a\}
  \enum
}
can be considered to be the kernel of a linear operator acting
on link fields with
values in $\Lie$. In particular,
\equation{
  \theta^a_{x,\mu}=\sum_{y,\nu}[\transs(V)](x,\mu;y,\nu)^{ab}
  \hat{\theta}^b_{y,\nu}.
  \enum
}
Since the functional integration measure $\rmD[U]$
is proportional to the maximal product of these
1-forms, it follows that
\equation{
  \rmD[U]=\rmD[V]\det\transs(V).
  \enum
}
The Jacobian of the map (1.1) is thus $\det\transs(V)$.

If the transformation satisfies
\equation{
  S(\trans(V))-\ln\det\transs(V)=\hbox{constant},
  \enum
}
the substitution $U\to V$ of the integration variables
in the functional integral 
maps the theory to the trivial one where the link variables
are completely decoupled from one another.
The expectation values (2.1) are then given by
\equation{
  \langle{\cal O}\rangle=
  \int\rmD[V]\,{\cal O}(\trans(V)).
  \enum
}
Such trivializing maps thus contain the entire 
dynamics of the theory.

Although the remark is likely to remain an academic one,
an intriguing observation is that the integral (2.9)
can be simulated simply by generating 
uniformly distributed random gauge fields. Subsequent
field configurations are uncorrelated in this case
and there are therefore no autocorrelations in
the data series for the observables $\cal O$.

\subsection 2.4 Transformation behaviour of the HMC algorithm

The HMC algorithm [\ref{HMC}] operates on the phase space
associated to the field manifold.
In particular, the transition $V\to V'$ in fig.~1 requires 
the equations of motion derived from the Hamilton function
\equation{
  \hat{H}(\hat{\mom},V)=
  \frac{1}{2}(\hat{\mom},\hat{\mom})+
  S(\trans(V))-\ln\det\transs(V)
  \enum
}
to be integrated, 
where $\hat{\mom}(x,\mu)\in\Lie$ is the canonical momentum of $V(x,\mu)$.

Although the complete update algorithm for the field $U$
described by fig.~1 looks different,
it is in fact equivalent to the HMC algorithm 
with a non-standard Hamilton function.
The equivalence can be established by noting
that the transformation from $V$ to $U$
preserves the symplectic 2-form
\equation{
  \hat{\Omega}=
  \sum_{x,\mu}\rmd\{\hat{\mom}^a(x,\mu)\hat{\theta}^a_{x,\mu}\},
  \enum
}
i.e.~$\Omega=\hat{\Omega}$,
if the momenta of the fields are transformed according to
\equation{
  \hat{\mom}^b(y,\nu)
  =\sum_{x,\mu}[\transs(V)](x,\mu;y,\nu)^{ab}\mom^a(x,\mu).
  \enum
}
The evolution of the transformed fields
is then governed by the Hamilton function
\equation{
  H(\mom,U)=
  \frac{1}{2}(\mom,K(U)\mom)+
  S(U)-\ln\det\transs(V),
  \enum
}
where
\equation{
  K(U)=\transs(V)^T\transs(V),\qquad V=\trans^{-1}(U).
  \enum
}
Note that the Jacobian in eq.~(2.13)
cancels when the momenta are integrated over in the 
functional integral.
The algorithm outlined in fig.~1 thus amounts
to applying the HMC algorithm with
a modified Hamilton function 
of the kind considered long ago by 
Duane et al.~[\ref{DuaneEtAl}].

\section 3. Transformations generated by flow equations

Flows in field space build up field transformations
from infinitesimal transformations.
The latter are generally easier to work with than
integral transformations, because they refer to the current
field only. Moreover, the differentiability and invertibility 
of the generated transformations is 
automatically guaranteed.

\subsection 3.1 Flows in field space

An infinitesimal field transformation,
\equation{
  U\to U+\eps Z(U)U+\rmO(\eps^2),
  \enum
}
is generated by a link field $[Z(U)](x,\mu)$ with
values in $\Lie$. The continuous composition of
such transformations
amounts to integrating a flow equation
\equation{
  \dot{U}_t=Z_t(U_t)U_t
  \enum
}
with respect to a ficticious time $t$. A simple choice
for the generator of the flow is
\equation{
  [Z_t(U)]^a(x,\mu)=\du^a_{x,\mu}\Wl_0(U),
  \enum
  \nexteq{2ex}
  \Wl_0(U)=\sum_{x,\mu\neq\nu}\,\tr\{U(x,\mu,\nu)\},
  \enum
}
where $U(x,\mu,\nu)$ denotes the plaquette loop in the
$(\mu,\nu)$-directions at the point $x$.
In the following, this flow will be referred to
as the ``Wilson flow''. 

If $Z_t(U)$ is a differentiable function of $t$ and $U$, the flow equation
(3.2) has a unique solution $U_t$ for any specified initial value $U_0=V$
and all $t\in(-\infty,\infty)$. Moreover,
the solution is differentiable with respect to
$t$ and $V$ (for a proof of these statements,
see ref.~[\ref{ArnoldI}], for example). 
It should be 
emphasized that the existence of the
solution for all times is non-trivial and can
only be guaranteed, without further
assumptions, because the field manifold is compact.

\subsection 3.2 Integrated transformations

At fixed $t$, the field $U_t$ is a well-defined
function of the initial field $V$. Through the
integration of the flow equation, one thus obtains a
differentiable transformation, $V\to U_t=\transt(V)$, of the
field space. The transformation is invertible and its
inverse is differentiable, because
the flow equation can be integrated backwards from
$t$ to $0$.
Moreover,
since the Jacobian $\det\transts(V)$
is equal to unity at $t=0$ and
does not pass through zero at any time, 
the transformation is also
orientation-preserving and thus fulfils all
requirements for an acceptable map of field space.

There is a useful compact expression for the Jacobian
which is obtained starting from the equations
\equation{
  {\rmd\over\rmd t}\ln\det\transts(V)=
  \Tr\{\dot{\trans}_{t,*}(V)\transts(V)^{-1}\},
  \enum
  \nexteq{3ex}
  [\dot{\trans}_{t,*}(V)](x,\mu;y,\nu)^{ab}
  =-2\,\tr\bigl\{\hat{\du}^b_{y,\nu}\{[Z_t(U_t)](x,\mu)U_t(x,\mu)\}
  U_t(x,\mu)^{-1}T^a
  \noenum
  \nexteq{2ex}
  {\phantom{[\dot{\trans}_{t,*}(V)](x,\mu;y,\nu)^{ab}=}}
  \kern1.9em-\hat{\du}^b_{y,\nu}U_t(x,\mu)U_t(x,\mu)^{-1}
  [Z_t(U_t)](x,\mu)T^a\bigr\}.
  \enum
}
Noting
\equation{
  \hat{\du}^b_{y,\nu}=\sum_{x,\mu}[\transs(V)](x,\mu;y,\nu)^{ab}
  \du^a_{x,\mu},
  \enum
}
a few lines of algebra then lead to the formula
\equation{
  \ln\det\transts(V)=\int_0^t\rmd s\,\sum_{x,\mu}
  \bigl\{\du^a_{x,\mu}[Z_s(U)]^a(x,\mu)\bigr\}_{U=U_s}.
  \enum
}
In the case of the Wilson flow, for example, the contribution
of the Jacobian to the action of the field $V$,
\equation{
  \ln\det\transts(V)=-\frac{16}{3}\int_0^t\rmd s\,\Wl_0(U_s),
  \enum
}
is proportional to the integral of
the Wilson plaquette action along the flow.

\section 4. Trivializing maps

Somewhat surprisingly, trivializing maps can, to some extent, 
be constructed expli\-cit\-ly in the pure gauge theory.
The construction is explained in this section, assuming
that the gauge action $S(U)$ is 
a sum of Wilson loops (plaquettes, rectangles, etc.).

\subsection 4.1 Trivializing flows

If the generator $Z_t(U)$ of the flow (3.2) is such that
\equation{
  \int_0^t\rmd s\,\sum_{x,\mu}
  \bigl\{\du^a_{x,\mu}[Z_s(U)]^a(x,\mu)\bigr\}_{U=U_s}
  =tS(U_t)+C_t,
  \enum
}
where $C_t$ may depend on $t$ but not on the fields,
the associated integrated transformations satisfy
\equation{
  S(\transt(V))-\ln\det\transts(V)=(1-t)S(\transt(V))-C_t.
  \enum
}
In particular, the transformation at $t=1$
is then a trivializing map.

Equation (4.1) is a rather implicit condition on the generator of
the flow. However, when differentiated with respect to 
$t$, it assumes a more tractable form,
\equation{
  \sum_{x,\mu}\bigl\{
  \du^a_{x,\mu}[Z_t(U)]^a(x,\mu)
  -t\du^a_{x,\mu}S(U)[Z_t(U)]^a(x,\mu)\bigr\}=
  S(U)+\dot{C_t},
  \enum
}
which involves the generator at time $t$ only.
Note that the differential condition (4.3) 
and the flow equation (3.2) imply eq.~(4.1),
i.e.~it suffices to find a generator $Z_t(U)$ that satisfies
eq.~(4.3).

\subsection 4.2 Existence of trivializing flows

Equation (4.3) is an inhomogeneous linear partial differential 
equation for the generator $Z_t(U)$.
Since it is a scalar equation,
one expects that there are many solutions.
In the following, the solution will be obtained in the form
\equation{
  [Z_t(U)]^a(x,\mu)=-\du^a_{x,\mu}\Sflow_t(U),
  \enum
}
where the action $\Sflow_t(U)$ is to be determined.

When inserted in eq.~(4.3), the ansatz (4.4) leads to the 
Laplace equation
\equation{
  \Lop\Sflow_t=S+\dot{C_t},
  \enum
  \nexteq{2ex}
  \Lop=
  \sum_{x,\mu}\bigl\{
  -\du^a_{x,\mu}\du^a_{x,\mu}
  +t\left(\du^a_{x,\mu}S\right)\du^a_{x,\mu}\bigr\}
  \enum
}
(for simplicity, the argument $U$ is now often omitted).
The operator $\Lop$ is elliptic and symmetric with respect to
the scalar product
\equation{
   (\phi,\psi)=\int\rmD[U]\,\rme^{-tS(U)}\phi(U)^*\psi(U).
   \enum
}
$\Lop$ has therefore a complete set of differentiable
eigenfunctions and a purely discrete spectrum
with no accumulation points (see ref.~[\ref{Gilkey}], sect.~1.6,
for example).
Moreover, since
\equation{
   (\phi,\Lop\phi)=\sum_{x,\mu}\bigl(\du^a_{x,\mu}\phi,\du^a_{x,\mu}\phi
   \bigr)\geq0,
   \enum
}
the function $\phi(U)=1$ is the only zero mode of $\Lop$
and all other eigenfunctions have eigenvalues separated
from the origin by a strictly positive spectral gap.

Now if one chooses $C_t$ to be such that
\equation{
   \dot{C}_t=-(1,S)/(1,1),
   \enum
}
the zero-mode component is removed from 
the right-hand side of eq.~(4.5) and 
\equation{
   \Sflow_t=\Lop^{-1}(S+\dot{C}_t).
   \enum
}
is then a well-defined expression that solves the equation.
The differentiability of $\Sflow_t$ with respect to $t$ and $U$
essentially follows from the ellipticity of $\Lop$ 
(appendix E).
A constructive proof of the existence of 
trivializing flows has thus been given.

\subsection 4.3 Expansion in powers of $t$

The solution (4.10) is well defined but still quite
implicit since it involves the inverse of an operator
acting on functions of the gauge field. In this and
the next subsection,
the solution is worked out analytically
in powers of $t$.

When the series
\equation{
  \Sflow_t=\sum_{k=0}^{\infty}t^k\Sflow^{(k)}
  \enum
}
is inserted in eq.~(4.5),
the matching of the terms of a given order in $t$ leads to the 
recursion
\equation{
  \Lap\Sflow^{(0)}=S+\dot{C}^{(0)},
  \enum
  \nexteq{2ex}
  \Lap\Sflow^{(k)}=
  -\sum_{x,\mu}\du^a_{x,\mu}S\,\du^a_{x,\mu}\Sflow^{(k-1)}
  +\dot{C}^{(k)},\qquad
  k=1,2,\ldots,
  \enum
}
for the actions $\Sflow^{(k)}$.
The Laplacian $\Lap$ coincides with the colour-electric
part of the Hamilton operator in lattice gauge theory
in $4+1$ dimensions.
In particular, its eigenfunctions are products
of $\Group$ representation
functions of the link variables.
Sums of
Wilson loops and products of Wilson loops, for example,
are eigenfunctions of $\Lap$
or can easily (algebraically) be
decomposed into eigenfunctions.

The solution of the recursion,
\equation{
  \Sflow^{(0)}=\Lap^{-1}S,
  \enum
  \nexteq{2ex}
  \Sflow^{(k)}=-\Lap^{-1}
  \sum_{x,\mu}\du^a_{x,\mu}S\,\du^a_{x,\mu}
  \Sflow^{(k-1)},\qquad k=1,2,\ldots,
  \enum
}  
is thus obtained in the form of sums of 
Wilson loops and products of Wilson loops.
Note that, 
as already mentioned in subsect.~4.2, 
the constant function is the only zero mode of $\Lap$.
Since the smallest non-zero eigenvalue
of $\Lap$ is $\frac{4}{3}$,
the right-hand sides of
eqs.~(4.14),(4.15) are therefore unambiguously determined
up to an irrelevant additive constant.

\subsection 4.4 Calculation of\/ $\Sflow^{(0)}$ and\/ $\Sflow^{(1)}$ 
                in the Wilson theory

For illustration, the first two terms of the 
series (4.11)
are now worked out explicitly for the plaquette action
[\ref{Wilson}]
\equation{
  \Sw(U)=-\frac{1}{6}\beta\,\Wl_0(U)
  \enum
}
(where $\beta=6/g_0^2$ denotes the inverse gauge coupling).
A short calculation,
using the completeness relation (A.5),
shows that the leading term is 
\equation{
  \Sflow^{(0)}=-\frac{1}{32}\beta\Wl_0=\frac{3}{16}\Sw.
  \enum
}
To this order and up to 
a rescaling of the time parameter $t$, the trivializing
flow in the Wilson theory thus coincides with the Wilson flow.

\topinsert
\vbox{
\vskip0.0cm
\centerline{\epsfxsize=8.5cm\epsfbox{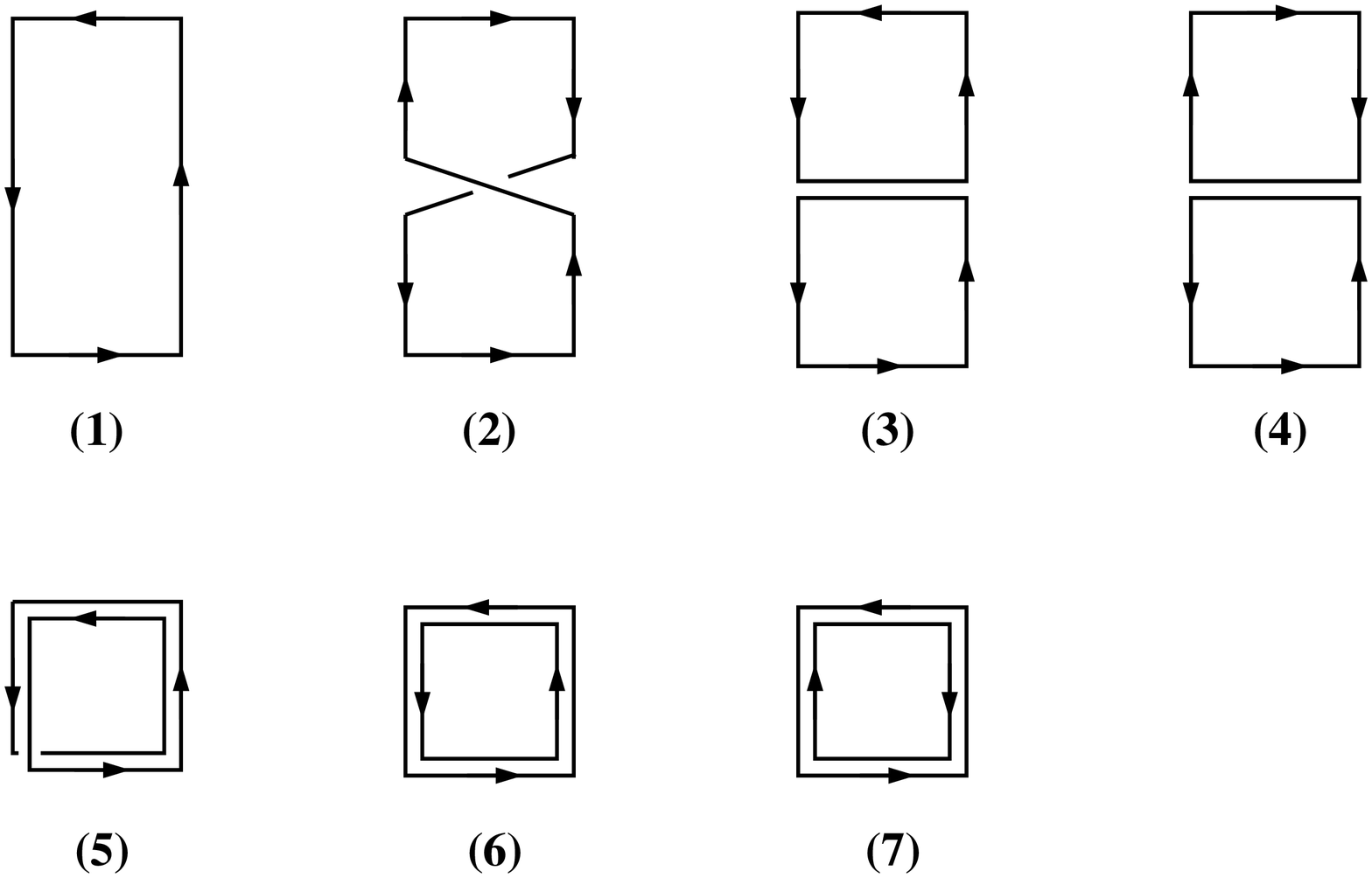}}
\vskip0.4cm
\figurecaption{%
Classes of loops and pairs of loops contributing to $\Sflow^{(1)}$
in the Wilson theory.
The loops $5-7$
reside on a single plaquette of the lattice. 
All other loops occupy two 
plaquettes which can lie in a plane or be at right angles
in three dimensions. 
}
\vskip0.0cm
}
\endinsert

The expression to be worked out at the next order is
\equation{
  \Sflow^{(1)}=
  -\frac{1}{192}\beta^2\Lap^{-1}\sum_{x,\mu}\du^a_{x,\mu}\Wl_0\,
  \du^a_{x,\mu}\Wl_0.
  \enum
}
The Wilson loops and products of Wilson loops
that can occur at this point derive from the
contractions of two plaquette loops with a common link.
Altogether there are then seven classes ${\cal C}_i$, $i=1,\ldots,7$, of
loops and pairs of loops to consider (see fig.~2). 
By summing the traces of the associated Wilson loops,
each class ${\cal C}_i$ defines an action
\equation{
  \Wl_i=\sum_{C\in{\cal C}_i}\tr\{U(C)\}
  \quad\hbox{if}\quad i=1,2,5,
  \enum
  \nexteq{2ex}
  \Wl_i=\sum_{\{C,C'\}\in{\cal C}_i}\tr\{U(C)\}\tr\{U(C')\}
  \quad\hbox{if}\quad i=3,4,6,7,
  \enum
}  
where $U(C)$ denotes the ordered product of the link variables
along the loop $C$. The sums in these equations extend over all possible 
positions of the loops on the lattice. Loops with opposite
orientation are considered to be different and are both included
in the sums.

Using the identity (A.5) again, some algebra now yields
\equation{
  \sum_{x,\mu}\du^a_{x,\mu}\Wl_0\,\du^a_{x,\mu}\Wl_0
  =\Wl_1-\Wl_2-\frac{1}{3}\Wl_3+\frac{1}{3}\Wl_4
  -2\Wl_5+\frac{2}{3}\Wl_6-\frac{4}{3}\Wl_7
  \enum
}
up to an additive constant. Furthermore,
\equation{
  \Lap\Wl_1=8\Wl_1,
  \enum
  \nexteq{2ex}
  \Lap\Wl_2=\frac{31}{3}\Wl_2+\Wl_4,
  \enum
  \nexteq{2ex}
  \Lap\Wl_3=11\Wl_3-\Wl_1,
  \enum
  \nexteq{2ex}
  \Lap\Wl_4=\frac{31}{3}\Wl_4+\Wl_2,
  \enum
  \nexteq{2ex}
  \Lap\Wl_5=\frac{28}{3}\Wl_5+4\Wl_6,
  \enum
  \nexteq{2ex}
  \Lap\Wl_6=\frac{28}{3}\Wl_6+4\Wl_5,
  \enum
  \nexteq{2ex}
  \Lap\Wl_7=12\Wl_7+\hbox{constant}.
  \enum
}
In the subspace of these functions,
the operator $\Lap$ can be easily inverted 
and the result
\equation{
  \Sflow^{(1)}=\frac{1}{192}\beta^2\left\{
  -\frac{4}{33}\Wl_1
  +\frac{12}{119}\Wl_2
  +\frac{1}{33}\Wl_3
  -\frac{5}{119}\Wl_4
  +\frac{3}{10}\Wl_5
  \right.
  \noenum
  \nexteq{2.0ex}
  {\phantom{\Sflow^{(1)}=\frac{1}{192}\beta^2\{}}
  \left.
  -\frac{1}{5}\Wl_6
  +\frac{1}{9}\Wl_7\right\}
  \enum
}
is thus obtained.
Note that the smallness of the numerical coefficients in this formula
is balanced, to some extent,
by the number of loops per lattice point
in the classes ${\cal C}_i$ (which are equal to 
$120$, $12$ and $6$ for $i=1,\ldots,4$, $i=5,6$ and $i=7$ respectively).

\subsection 4.5 Miscellaneous remarks

\noindent
(a) {\it Higher orders}.
The actions $\Sflow^{(k)}$, $k\geq2$, can 
be computed algebraically
following the steps taken in the previous subsection.
Since all loops generated by contracting
a plaquette loop with the loops at order $k-1$
must be considered, the work required for 
the calculation 
is however rapidly growing with $k$.

\vskip0.5ex
\noindent
(b) {\it Locality and convergence}. The series (4.11) is an
expansion in local terms whose footprint on the lattice 
increases proportionally to the order $k$. 
At the values of $t$, where the expansion converges, 
the action $\Sflow_t$ is then guaranteed to be local as well.

The norm estimates in appendix E imply a lower bound on
the convergence radius of the series, but this bound
is rather poor and vanishes in the infinite-volume limit.
It seems nevertheless plausible that 
the series has a non-zero convergence radius
in this limit,
because the inverse of the operator $\Lop$ 
in eq.~(4.10) is likely to remain bounded
in a complex neighbourhood of $t=0$.
An analysis that
takes the locality properties of $\Lop$
into account will however be required to show this.

\vskip0.5ex
\noindent
(c) {\it Truncation of the expansion}.
If all terms in the series (4.11) of order $k\geq n$ are dropped,
one obtains an approximately trivializing flow that 
satisfies eq.~(4.2) up to an additive correction
of order $t^{n+1}$. 
An at least partial cancellation of the action is
achieved in this case as long as $t$ is
sufficiently small for the correction
to be strongly suppressed.

\vskip0.5ex
\noindent
(d) {\it Smoothing property}. 
The Wilson flow satisfies
\equation{
  {\rmd\over\rmd t}\Sw(U_t)=-\frac{3}{16}\sum_{x,\mu}
  \bigl\{\du^a_{x,\mu}\Sw(U)
  \du^a_{x,\mu}\Sw(U)\bigr\}_{U=U_t}\leq 0
  \enum
}
and therefore lowers the Wilson action as $t$ increases.
To leading order in $t$, the trivializing flow 
constructed in this section for the Wilson theory thus 
has a smoothing effect on the gauge field.
On the other hand, if the flow is followed in the reverse
direction, the gauge field tends to become rougher.

\vskip0.5ex
\noindent
(e) {\it Topological charge sectors}.
In lattice QCD, the topological (instanton) 
sectors are not a property of the field manifold alone, but
are expected to emerge dynamically when the continuum limit is approached.
The fact that trivializing maps completely ``straighten out'' 
the sectors is therefore not in conflict with the topological
properties of the field space.

\vskip0.5ex
\noindent
(f) {\it Renormalization group}. 
By composing the trivializing map $U=\trans_1(V)$
in the Wilson theory 
with its inverse at another value of the gauge coupling,
one obtains a group of transformations whose only effect 
on the action is a shift of the coupling. 
The locality properties of these transformations are
not transparent, however, and could be quite different from the ones
of a Wilsonian ``block spin'' transformation.

\section 5. Numerical integration of the Wilson flow

The discussion in sect.~1 now suggests 
to combine the HMC algorithm with 
the field transformations generated by the 
trivializing flow constructed in the previous section.
In particular, if the Wilson gauge action is used,
the transformations generated by the Wilson flow
may lead to an algorithm with improved sampling efficiency.

\subsection 5.1 Forward integration

There is a wide range of numerical integration methods 
that can in principle be used
to integrate the Wilson flow 
(see ref.~[\ref{HairerEtAl}], for example). 
The Euler scheme discussed in the following
performs the integration in time steps of size $\eps$ and 
updates the link variables one after another 
according to
\equation{
   U(x,\mu)\to U'(x,\mu)=\rme^{\eps [Z(U)](x,\mu)}U(x,\mu),
   \enum
}
where
\equation{
   [Z(U)](x,\mu)=T^a\du^a_{x,\mu}\Wl_0(U).
   \enum
}
Starting from the gauge field $U_t$ at time $t$, 
the field at time $t+\eps$ 
is thus obtained by 
running through all links $(x,\mu)$ on the lattice 
and updating the link variable residing there.
Note that the ordering of the links matters, since
the old value of $U(x,\mu)$ is replaced by the 
new one before going to the next link.

The generator of the flow is explicitly given by
\equation{
   [Z(U)](x,\mu)=
   -\sum_{\nu\neq\mu}\Proj\bigl\{
   U(x,\mu)U(x+\hat{\mu},\nu)U(x+\hat{\nu},\mu)^{-1}U(x,\nu)^{-1}
   \noenum
   \nexteq{1.5ex}
   {\phantom{[Z_t(U)](x,\mu)=}}
   +U(x,\mu)U(x+\hat{\mu}-\hat{\nu},\nu)^{-1}U(x-\hat{\nu},\mu)^{-1}
   U(x-\hat{\nu},\nu)\bigr\},   
   \enum
}
where $\hat{\mu}$ denotes the unit vector in direction $\mu$ and
\equation{
   \Proj\{M\}=\frac{1}{2}\bigl(M-M^{\dagger}\bigr)-
   \frac{1}{6}\tr\bigl(M-M^{\dagger}\bigr)
   \enum
}
projects any $3\times3$ matrix $M$ to $\Lie$.
The Euler integration of the Wilson flow thus
amounts to applying a number of ``stout smearing'' steps [\ref{Stout}],
except that the link variables are here updated one by one rather than
all at once.

\subsection 5.2 Backward integration

The application of $n$ Euler sweeps
maps the initial field 
$V=U_0$ to the field $U=U_{n\eps}$ at time $t=n\eps$. If $t$ is held
fixed and $n$ is taken to infinity, this map converges to the 
transformation obtained by integrating the Wilson flow exactly.
However, the HMC algorithm may potentially be combined
with the map defined by the Euler integrator at fixed $n$ and $\eps$.
For this proposition to be a viable option,
the transformation must be invertible, i.e.~one must
be able to trace back the Euler integration
by inverting the link
update steps one by one in the reverse order.

The question is thus whether 
eq.~(5.1) has a unique solution $U(x,\mu)$ given $U'(x,\mu)$ 
(and keeping all other link variables fixed).
As explained in appendix D, 
the answer is affirmative, 
for arbitrary values of the field variables, if
\equation{
  |\eps|<\frac{1}{8}.
  \enum
}
Moreover, in this range of $\eps$,
the solution $U(x,\mu)=\rme^{-\eps X_{*}}U'(x,\mu)$
can be obtained through the fixed-point iteration
\equation{
  X_0=0,
  \enum
  \nexteq{2ex}
  X_{n+1}=\bigl\{[Z(U)](x,\mu)\bigr\}_{
  U(x,\mu)=\rme^{-\eps X_n}U'(x,\mu)},
  \qquad
  n=0,1,2,\ldots,
  \enum
}
which converges to $X_{*}$ at an exponential rate.

In appendix D it is also shown that the 
Jacobian of the transformation (5.1)
is strictly positive in the range (5.5).
The field transformations obtained through the Euler
integration of the Wilson flow are therefore orientation-preserving
diffeomorphisms of the field manifold, as required
for acceptable maps of field space.

\subsection 5.3 Jacobian matrix of the Euler integrator

The Euler step (5.1) amounts to applying the field transformation
\equation{
  [\euler{x,\mu}(U)](y,\nu)=\cases{
  \rme^{\eps [Z(U)](x,\mu)}U(x,\mu)
  & if $(y,\nu)=(x,\mu)$,\cr
  \noalign{\vskip1.5ex}
  U(y,\nu) & otherwise,\cr}
  \enum
}
to the current gauge field. An Euler sweep is then the composition
product of these transformations over all links $(x,\mu)$.
It is straightforward to show that the Jacobian matrix of
a composed transformation is the product of the Jacobian matrices of 
the factors. The Jacobian matrix of the 
Euler integrator is therefore
an ordered product of the Jacobian matrices 
\equation{
  [\euler{x,\mu,*}(U)](y,\nu;z,\rho)^{ac}=
  -2\,\tr\bigl\{\du^c_{z,\rho}[\euler{x,\mu}(U)](y,\nu)
  [\euler{x,\mu}(U)](y,\nu)^{-1}T^a\bigr\}
  \enum
}
of the one-link transformations (5.8).

The matrix (5.9) can be expressed through the derivative
\equation{
  [Z_{*}(U)](y,\nu;z,\rho)^{bc}=
  \du^c_{z,\rho}[Z(U)]^b(y,\nu)
  \enum
}
of the generator of the Wilson flow.
Explicitly, one finds that
\equation{
  [\euler{x,\mu,*}(U)](y,\nu;z,\rho)^{ac}=
  \delta^{ac}\delta_{yz}\delta_{\nu\rho}
  +\delta_{xy}\delta_{\mu\nu}
  \Bigl\{\left(\rme^{\Ad X}-1\right)^{ac}\delta_{yz}\delta_{\nu\rho}
  \noenum
  \nexteq{2.5ex}
  {\phantom{[\euler{x,\mu,*}(U)](y,\nu;z,\rho)^{bc}=}}
  +\eps J(-X)^{ab}[Z_{*}(U)](y,\nu;z,\rho)^{bc}\Bigr\}_{X=\eps[Z(U)](x,\mu)},
  \enum
}
where use was made of the $\Group$ formulae 
listed in appendix A and B.

\subsection 5.4 Jacobian of the integrated transformations

The Euler integrator generates a sequence of fields
\equation{
  V=U_0\to U_{\eps}\to U_{2\eps}\to\ldots\to U_{n\eps}=U
  \enum
}
by sweeping through the lattice $n$ times and 
updating the link variables one by one in a specified order.
In the following, the intermediate field configurations 
obtained starting from $U_{k\eps}$ 
and updating the link variables 
on all links that come before $(x,\mu)$ will be denoted by
$U_{k\eps,[x,\mu]}$. 
In particular, $U_{k\eps,[x,\mu]}=U_{k\eps}$ if $(x,\mu)$ is the first link
and 
\equation{
  U_{k\eps,[y,\nu]}=\euler{x,\mu}(U_{k\eps,[x,\mu]})
  \enum
}
if $(y,\nu)$ follows $(x,\mu)$ in the chosen link order.

Since the transformation $V\to U=\transne(V)$
is a composition product of one-link update steps,
its Jacobian
\equation{
  \det\transnes(V)=
  \prod_{k=0}^{n-1}
  \prod_{x,\mu}\det\euler{x,\mu,*}(U_{k\eps,[x,\mu]})
  \enum
}
factorizes into the product of the Jacobians of the steps.
The latter coincide with the determinants of certain real 
$8\times8$ matrices given explicitly
in appendix C.

\section 6. Proposed simulation algorithm for lattice QCD

With respect to the QCD simulation algorithms used to date,
the combination of the transformations obtained
through the Euler integration of the Wilson flow
and the HMC algorithm is expected to 
sample the topological sectors more quickly
and to be generally more efficient.
The use of the Wilson flow is suggested
if the gauge action coincides with the Wilson 
plaquette action. For other actions,
the appropriate flow can be easily constructed
following the lines of sect.~4.

\subsection 6.1 Choice of the parameters

The parameters of the Euler integrator
are the integration step size $\eps$ and the number $n$ of 
Euler sweeps that are applied. One also needs to choose
a definite ordering of the links of the lattice.

The step size $\eps$ should be positive 
and not larger than, say, $1/16$ so that
the in\-ver\-ti\-bi\-lity of the 
Euler integrator is guaranteed within a safe margin.
Some tuning of the integration time $n\eps$
will certainly be required in order to maximize
the efficiency of the algorithm.
Note that the unit of time differs 
from the one used in subsect.~4.4, i.e.~setting $t=1$ there
corresponds to 
an integration time $n\eps=\beta/32$.

The ordering of the links can be chosen arbitrarily.
One may, for example, first visit the links $(x,0)$
on all even points $x$, then the links $(x,0)$ on 
all odd points, then the links $(x,1)$ on all even points,
and so on.
This ordering is well suited for parallel processing,
since the link variables in a given direction on the even (odd) sites
are decoupled from one another and can therefore be updated in parallel.

\subsection 6.2 Force calculation

At the beginning of an update cycle,
the current gauge field
$U$ is transformed to the field $V=\transne^{-1}(U)$ by applying 
$n$ backward Euler sweeps to $U$. 
The force that drives the molecular-dynamics evolution of $V$
is then given by
\equation{
  F(z,\rho)^c=\hat{\du}^c_{z,\rho}\left\{
  S(\transne(V))-\ln\det\transnes(V)\right\},
  \enum
}
where the action $S(U)$ now includes the usual sea-quark pseudo-fermion actions
(for simplicity, the dependence on the
pseudo-fermion fields is suppressed).

\topinsert
\vbox{
\vskip0.0cm
\centerline{\epsfxsize=3.6cm\epsfbox{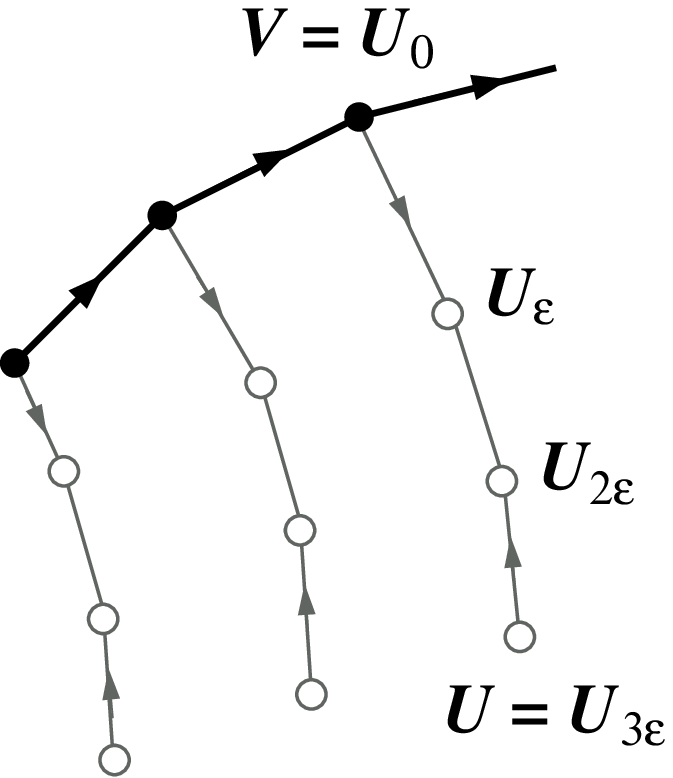}}
\vskip0.4cm
\figurecaption{%
The proposed algorithm evolves the field $V$ using the standard
HMC algorithm (thick line). The force that drives the 
molecular-dynamics evolution
is obtained by forward integration of the Wilson flow 
and subsequent backward propagation of  
the derivatives of the action and the Jacobian
of the flow ($n=3$ in this figure). 
}
\vskip0.0cm
}
\endinsert

Each time the force is to be calculated, 
the current field $V$ must be transformed 
to $U$ again by applying $n$ 
forward Euler sweeps (see fig.~3). The fields 
$U_0,U_{\eps},\ldots,U_{n\eps}$ generated in this process should
be stored in memory so that they will be available
when the force is propagated from $U$ to $V$.
Note that the intermediate fields 
\equation{
  U_{k\eps,[x,\mu]}(y,\nu)=\cases{
  U_{k\eps}(y,\nu) & if $(y,\nu)\geq(x,\mu)$,\cr
  \noalign{\vskip1ex}
  U_{k\eps+\eps}(y,\nu) & otherwise,\cr}
  \enum
}
are then also available.

The factorization (5.14) of the Jacobian implies
\equation{
   F(z,\rho)^c=\hat{\du}^c_{z,\rho}S(U_{n\eps})
   -\sum_{k=0}^{n-1}\sum_{x,\mu}
  \hat{\du}^c_{z,\rho}\ln\det\euler{x,\mu,*}(U_{k\eps,[x,\mu]}).
  \enum
}
Moreover, 
\equation{
  \hat{\du}^c_{z,\rho}S(U_{n\eps})=
  \sum_{y,\nu}\left.\du^a_{y,\nu}S(U)\right|_{U=U_{n\eps}}
  \transnes(y,\nu;z,\rho)^{ac},
  \enum
}
and there is a similar formula for the other terms in eq.~(6.3) 
involving the Jacobian matrices of the 
transformations $V\to U_{k\eps,[x,\mu]}$.
All these matrices, as well as the one in eq.~(6.4),
are products of the Jacobian matrices of the one-link 
transformations (5.8).
The force can therefore be computed recursively 
as follows:

{
\parindent=1.5em
\vskip1.5ex
\item{1.} Set $U=U_{n\eps}$ and $F(z,\rho)^c=\du^c_{z,\rho}S(U)$.

\vskip1.0ex
\item{2.} For $k$ from $n-1$ to $0$, 
run through all links $(x,\mu)$ in reverse order,
set $U=U_{k\eps,[x,\mu]}$
and update the force according to
\equation{
   F(z,\rho)^c\to\sum_{y,\nu}F(y,\nu)^a
   [\euler{x,\mu,*}(U)](y,\nu;z,\rho)^{ac}
   -\du^c_{z,\rho}\ln\det\euler{x,\mu,*}(U).
   \enum
}}

\vskip-1.5ex
\noindent
Note that the field $U$ backtracks
the forward integration of the Wilson flow
in the course of the recursion.
Since the Jacobian matrix $\euler{x,\mu,*}(U)$ differs from unity
only on the links sharing a plaquette with $(x,\mu)$,
the total computational effort required 
for the propagation (6.5) of the force
is expected to be similar to the one required for
$n$ applications of a nearest-neighbour
gauge-covariant difference operator to the force field.

\subsection 6.3 Domain-decomposed algorithm

Field transformations 
can also be combined with the DD-HMC algorithm [\ref{DDHMC}].
Only the so-called active link variables
are transformed in this case, but the transformations may depend on 
the inactive field components.

In the pure gauge theory, it is then again possible to 
construct trivializing flows. They operate
on the active link variables and
contract the gauge action in each do\-main to a constant
(i.e.~an expression depending on the inactive link variables
only).
These flows are not the same as the ones constructed
in sect.~4, but can be obtained by solving
the differential equations derived there.
In particular,
the trivializing flow
for the Wilson action is, to leading order, 
a slightly modified Wilson flow,
where the plaquettes containing $\nu$ active links 
are given the weight $4/\nu$.

\section 7. Concluding remarks

Trivializing maps and flows in
field space have here been discussed with a particular application
in mind. 
The underlying concepts are fairly general, however, and
may have other uses in rigorous constructive work,
numerical perturbation theory or in connection with 
renormalizable smoothing techniques, for example.

Whether the proposed combination of the Wilson flow
and the HMC algorithm does indeed 
sample the topological sectors in lattice QCD more efficiently
than the simulation algorithms used so far remains to be determined.
As the lattice spacing is taken to smaller and smaller values,
the quark fields
may eventually have to be included in the flow and perhaps also
the next-to-leading order correction
discussed in sect.~4. Trivializing maps
in the presence of matter fields is, in any case,
a subject that deserves to be
studied in its own right.

\vskip0.3ex
In the course of this work, I profited from 
many discussions with Filippo Palombi and Stefan Schaefer 
of various questions related to 
the slow topology-switching in current lattice QCD simulations.
I also wish to thank
Stefan Schaefer, Rainer Sommer and Francesco Virotta for
sharing some of their simulation results before publication.

\appendix A. SU(3) notation 

\vskip-2.5ex 

\subsection A.1 Group generators

The Lie algebra $\Lie$ of $\Group$ 
may be identified with the space of all anti-hermitian
traceless $3\times3$ matrices.
With respect to a 
basis $T^a$, $a=1,\ldots,8$, of such matrices, 
the elements $X\in\Lie$ are given by
\equation{
  X=X^aT^a,
  \enum
}
where $(X^1,\ldots,X^8)\in\rz^8$
(repeated group indices are automatically summed over).

The generators $T^a$ are assumed to satisfy
the normalization condition
\equation{
  \tr\{T^aT^b\}=-\frac{1}{2}\delta^{ab}.
  \enum
}
The structure of the Lie algebra is then encoded in the commutators
\equation{
  [T^a,T^b]=f^{abc}T^c,
  \enum
}
while the completeness of the generators implies
\equation{
  \{T^a,T^b\}=-\frac{1}{3}\delta^{ab}+id^{abc}T^c,
  \enum
  \nexteq{2ex}
  T^a_{\alpha\beta}T^a_{\gamma\delta}=
  -\frac{1}{2}\left\{
  \delta_{\alpha\delta}\delta_{\beta\gamma}-\frac{1}{3}\delta_{\alpha\beta}
  \delta_{\gamma\delta}\right\}.
  \enum
}
It follows from these equations that the structure constants
$f^{abc}$ and the tensor $d^{abc}$ are both real.
Moreover, $f^{abc}$
is totally anti-symmetric in the indices and $d^{abc}$ totally symmetric
and traceless.

\subsection A.2 Adjoint representation

The representation space of the adjoint representation of $\Lie$
is the Lie algebra itself, 
i.e.~the elements $X$ of $\Lie$ are represented by linear 
transformations
\equation{
  \Ad X:\,\Lie\mapsto\Lie,
  \enum
  \nexteq{2ex}
  \Ad X\cdot Y=[X,Y]\quad\hbox{for all}\quad Y\in\Lie.
  \enum
}
The action of $\Ad X$ on the group generators
is given by
\equation{
  \Ad X\cdot T^b=T^a(\Ad X)^{ab},
  \enum
}
where
\equation{
  (\Ad X)^{ab}=-f^{abc}X^c
  \enum
}
is a real antisymmetric $8\times 8$ matrix.

\subsection A.3 Matrix norms

The natural scalar product in $\Lie$ is
\equation{
  (X,Y)=X^aY^a=-2\,\tr\{XY\}.
  \enum
}
In particular, $\|X\|=(X,X)^{1/2}$ is a possible definition of 
the norm of $X\in\Lie$.

Another useful matrix norm derives from the square norm
\equation{
  \|v\|_2=\{|v_1|^2+|v_2|^2+|v_3|^2\}^{1/2}
  \enum
}
of complex colour vectors $v$. If $A$ is any complex $3\times3$ matrix,
one defines
\equation{
  \|A\|_2=\max_{\|v\|_2=1}\|Av\|_2.
  \enum
}
This norm satisfies
\equation{
   \|A+B\|_2\leq\|A\|_2+\|B\|_2,
   \qquad
   \|AB\|_2\leq\|A\|_2\|B\|_2,
   \enum
}
for all matrices $A,B$. Moreover, if $A$ is hermitian or
antihermitian, $\|A\|_2$ is equal to the maximum of the absolute
values of its eigenvalues.

\appendix B. Properties of the SU(3) exponential function

\vskip-2.5ex

\subsection B.1 Lipschitz bound

For any $X,Y\in\Lie$, the relation
\equation{
  \|\rme^X-\rme^Y\|_2=
  \|1-\rme^{-X}\rme^Y\|_2
  \enum
}
follows from the fact that $\rme^X$ is unitary.
Using the identity
\equation{
  1-\rme^{-X}\rme^Y=
  \int_0^1\rmd s\,\rme^{-sX}(X-Y)\rme^{sY}
  \enum
}
and the subadditivity (A.13) of the norm,
the Lipschitz bound 
\equation{
  \|\rme^X-\rme^Y\|_2\leq\|X-Y\|_2
  \enum
}
is then obtained.

\subsection B.2 Differential of the exponential map

Let $X$ be an element of the Lie algebra $\Lie$. 
A linear mapping $J(X):\Lie\mapsto\Lie$ is then defined by
\equation{
  J(X)\cdot Y=\rme^{-X}\left.{\rmd\over\rmd t}\rme^{X+tY}\right|_{t=0}
  \quad\hbox{for all $Y\in\Lie$.}
  \enum
}
$J(X)$ is referred to as the differential of the exponential map.
Scaling the expo\-nents by a parameter $s$, as above,
one obtains the representation
\equation{
  J(X)\cdot Y=
  \int_0^1\rmd s\,\rme^{-sX}Y\rme^{sX}
  \enum
}
and thus the expansion
\equation{
  J(X)=1+\sum_{k=1}^{\infty}{(-1)^k\over(k+1)!}(\Ad X)^k
  \enum
}
which is absolutely convergent for any $X\in\Lie$.

The action of $J(X)$ on the group generators $T^a$
is given by
\equation{
  J(X)\cdot T^b=T^aJ(X)^{ab},
  \enum
}
where $J(X)^{ab}$ is a real $8\times8$ matrix.
Note that 
\equation{
  J(X)^T=J(-X)=\rme^{\Ad X}J(X),
  \qquad
  J(X)\Ad X=1-\rme^{-\Ad X}.
  \enum
}
Moreover, eq.~(B.5) implies
\equation{
  \|J(X)\cdot Y\|_2\leq\|Y\|_2
  \enum
}
for all $X,Y\in\Lie$.

\appendix C. Jacobian of the Euler step

The Jacobian matrix (5.11) of the Euler step
is equal to unity except for some non-zero elements
along the row $(y,\nu)=(x,\mu)$. Its determinant
therefore coincides with the determinant of the $(x,\mu;x,\mu)$-element
\equation{
  A^{ac}=\left\{\left(\rme^{\Ad X}\right)^{ac}  
  +\eps J(-X)^{ab}[Z_{*}(U)](x,\mu;x,\mu)^{bc}
  \right\}_{X=\eps[Z(U)](x,\mu)}
  \enum
}
of the matrix.

The derivative $[Z_{*}(U)](x,\mu;x,\mu)^{bc}$ 
can be worked out explicitly in terms of
the plaquette sum
\equation{
  M=\sum_{\nu\neq\mu}\bigl\{
  U(x,\mu)U(x+\hat{\mu},\nu)U(x+\hat{\nu},\mu)^{-1}U(x,\nu)^{-1}+
  \noenum
  \nexteq{1.0ex}
  {\phantom{M=\sum_{\nu\neq\mu}\bigl\{}}
  U(x,\mu)U(x+\hat{\mu}-\hat{\nu},\nu)^{-1}U(x-\hat{\nu},\mu)^{-1}
  U(x-\hat{\nu},\nu)\bigr\}.
  \enum
}
A few lines of algebra then lead to the expression
\equation{
  A^{ac}=B^{ac}+\frac{1}{2}\eps C^{ab}\Bigl\{
  id^{bcd}\tr\{T^d(M+M^{\dagger})\}
  -\frac{1}{3}\delta^{bc}\tr\{M+M^{\dagger}\}\Bigl\}
  \enum
}
where 
\equation{
  B=\frac{1}{2}\bigl(\rme^{\Ad X}+1\bigr),
  \enum
  \nexteq{2ex}
  C=J(-X),\qquad X=-\eps\Proj\{M\}.
  \enum
}
In particular,
\equation{
  \det A=1-\frac{4}{3}\eps\,\tr\{M+M^{\dagger}\}+\rmO(\eps^2),
  \enum
}
as expected from eq.~(3.9).

\vfill\eject

\appendix D. Inversion of the Euler step

\vskip-2.5ex

\subsection D.1 Basic norm bounds

For any complex $3\times3$ matrix $M$, 
the inequality
\equation{
  \|\Proj\{M\}\|_2\leq\frac{4}{3}\|M\|_2
  \enum
}
can be established in a few lines. One first observes that
\equation{
  \frac{1}{2}\bigl(M-M^{\dagger}\bigr)=A D A^{-1},
  \qquad A\in\Group,
  \enum
}
where $D$ is a diagonal matrix with diagonal elements 
$\lambda_1,\lambda_2,\lambda_3$. Setting
\equation{
  \bar{\lambda}=\frac{1}{3}\sum_{k=1}^3\lambda_k,
  \enum
}
the estimates
\equation{
  \|\Proj\{M\}\|_2=\max_k
  \left|\lambda_k-\bar{\lambda}\right|
  \noenum
  \nexteq{2ex}
  {\phantom{\|\Proj\{M\}\|_2}}
  \leq\frac{4}{3}\max_k\left|\lambda_k\right|=
  \frac{2}{3}\|M-M^{\dagger}\|_2\leq\frac{4}{3}\|M\|_2
  \enum
}
then show that the inequality (D.1) holds for all matrices $M$.

An immediate consequence of the bound (D.1) and the Lipschitz bound (B.3) 
is that
\equation{
  \bigl\|\Proj\bigr\{A\bigl(\rme^X-\rme^Y\bigr)B\bigr\}\bigr\|_2\leq
  \frac{4}{3}\|X-Y\|_2
  \enum
}
for all $A,B\in\Group$ and $X,Y\in\Lie$.

\subsection D.2 Solution of eq.~(5.1)

For a given link $(x,\mu)$ and any fixed $U'(x,\mu)\in\Group$, 
the function
\equation{
  f(X)=\bigl\{[Z(U)](x,\mu)\bigr\}_{
  U(x,\mu)=\rme^{-\eps X}U'(x,\mu)}
  \enum
}
maps $X\in\Lie$ back to $\Lie$. Recalling eq.~(5.3), the inequality
(D.5) immediately implies that
\equation{
  \|f(X)-f(Y)\|_2\leq k\|X-Y\|_2, 
  \qquad k=8|\eps|.
  \enum
}  
The function $f(X)$ is therefore a strict contraction if
the integration step size $\eps$ is in the range (5.5) (which is assumed
to be the case from now on).

It is not difficult to prove that
strict contractions in a complete metric space
have a unique fixed point
(see ref.~[\ref{ReedSimonI}], Theorem V.18, for example).
In the present case, the fixed point
$X_{*}$ can be computed by noting that 
the sequence $X_0=0,X_1,X_2,\ldots$ generated through
the recursion $X_{n+1}=f(X_n)$ satisfies
\equation{
  \|X_n-X_{*}\|_2\leq k\|X_{n-1}-X_{*}\|_2
  \enum
}
and therefore rapidly converges to $X_{*}$.
The matrix 
\equation{
  U(x,\mu)=\rme^{-\eps X_{*}}U'(x,\mu)
  \enum
}
then provides a solution of eq.~(5.1).
Moreover, there is no other solution,
because the fixed point $X_{*}$ of $f(X)$ is unique.

\subsection D.3 Positivity of the Jacobian

In order to prove that
the determinant of the Jacobian matrix (C.1) is positive
at all $\eps$ in the range (5.5),
it suffices to show that the matrix has no zero mode, i.e.~that 
\equation{
  A^{ac}Y^c=0
  \enum
}
implies $Y=0$. 
To this end, eq.~(D.10) is first written in the form
\equation{
  Y^a=-\eps J(X)^{ab}W^b,
  \enum
}
where 
\equation{
  X=\eps[Z(U)](x,\mu),\qquad
  W^b=[Z_{*}(U)](x,\mu;x,\mu)^{bc}Y^c.
  \enum
}
Recalling eq.~(5.10), the formula
\equation{
  W=\lim_{t\to0}{1\over t}\left\{
  \left.[Z(U)](x,\mu)\right|_{U(x,\mu)\to\rme^{tY}U(x,\mu)}-[Z(U)](x,\mu)
  \right\}
  \enum
}
may then be derived from which one infers that
\equation{
  \|W\|_2\leq 8\|Y\|_2.
  \enum
}
The inequality (D.5) has here been used again and also the fact that 
the norm is a continuous map from $\Lie$ to $\rz$.
The combination of eq.~(D.11) and the bounds (B.9) and (D.14)
now implies $\|Y\|_2\leq k\|Y\|_2$ and thus $Y=0$.

\appendix E. Differentiability of $\mib\Sflow_t$

The action $\Sflow_t$ solves an 
elliptic partial differential equation with smooth coefficients
and is therefore guaranteed (by elliptic regularity)
to be a differentiable function of the 
gauge field $U$. 
In this appendix, the simultaneous
differentiability in the time $t$
is established by expanding $\Sflow_t$
in powers of $t-t_0$ around any fixed time $t_0$.
Using Sobolev norms,
the expansion can be shown to converge
if $t$ is sufficiently close to $t_0$.
The pointwise
uniform convergence of the series and its derivatives,
and therefore the differentiability of $\Sflow_t$,
then follows from
Sobolev's lemma.

\subsection E.1 Sobolev spaces on the field manifold

The definition of the Sobolev spaces on a compact manifold
is quite involved and will not be reviewed here. An introduction
to the subject and the proof of all statements made in this subsection
is given in the first three sections of ref.~[\ref{Gilkey}],
for example.

Let $\Cinfty$ be the space of differentiable functions on the field
manifold and $H_k$ the associated Sobolev space of order $k\in\gz$. The latter
is the completion of $\Cinfty$ with respect to a certain norm $\|\cdot\|_k$.
A characteristic feature of these norms is that 
the bounds
\equation{
  \|\frak D\phi\|_k\leq c_k\|\phi\|_{k+p}
  \enum
}
hold for all $\phi\in\Cinfty$ and 
any differential operator $\frak D$ of order $p$
with smooth coefficients,
where the constants $c_k$ depend on $\frak D$ but not on $\phi$.
Such differential operators thus extend to
bounded linear operators from $H_{k+p}$ to $H_k$. 
Moreover, 
$\|\phi\|_k\leq\|\phi\|_l$ if $k<l$
and therefore 
$H_l\subset H_k$.

A fairly concrete description of the
Sobolev space $H_k$ can be given when $k$ is even 
and non-negative. For any $t\in\rz$ and $j=0,1,2,\ldots$,
another norm $\|\phi\|_{t,j}$ of $\phi\in\Cinfty$
may be defined through
\equation{
  \|\phi\|_{t,j}=\|\phi\|+\|\Lop^j\phi\|,
  \enum
}
where $\|\cdot\|$ is the norm associated to the scalar product (4.7).
The fact that $\Lop$ is a second-order elliptic differential operator
then implies
\equation{
  \|\phi\|_{2j}\leq a_{t,j}\|\phi\|_{t,j},
  \qquad 
  \|\phi\|_{t,j}\leq b_{t,j}\|\phi\|_{2j}
  \enum
}
for some constants $a_{t,j},b_{t,j}$. 
In particular,
$H_{2j}$ is the completion of $\Cinfty$ with respect to the 
norm $\|\cdot\|_{t,j}$.

\subsection E.2 Properties of the inverse of $\Lop$

Let $\Pop$ be the orthogonal projector to the zero mode of $\Lop$
in the Hilbert space with scalar product (4.7).
Its action on any function $\phi\in\Cinfty$ is given by
\equation{
  \Pop\phi=(1,\phi)/(1,1).
  \enum
}
Note that $\Pop$ projects $\phi$ to a constant, but the constant
depends on the definition of the scalar product $(\cdot\,,\cdot)$
and therefore on $t$.
The action of the inverse $\Gop$ of $\Lop$ on $\phi$
is then determined by the equations
\equation{
  \Lop\Gop\phi=(1-\Pop)\phi,
  \qquad
  \Pop\Gop\phi=0.
  \enum
}
As already mentioned in subsect.~4.2, the ellipticity of $\Lop$ 
and the absence of further zero modes
imply that these equations have, for any fixed $t$, 
a unique differentiable solution $\Gop\phi$. 
Evidently, $\Sflow_t=\Gop S$.

Since $\Lop$ has a spectral gap in the subspace orthogonal to 
the zero mode, $\Gop$ is a bounded operator with respect
to the norm $\|\cdot\|$.
As a consequence, there exists a constant $g_t$ such that
\equation{
  \|\Gop\phi\|_{t,j}\leq g_t\|\phi\|_{t,j-1}
  \enum
}
for all $j=1,2,\ldots$ and all $\phi\in\Cinfty$. Moreover,
recalling the inequalities (E.3), one concludes that
\equation{
  \|\Gop\phi\|_{2j}\leq g_{t,j}\|\phi\|_{2j-2}
  \enum
}
for some other constants $g_{t,j}$.

\subsection E.3 Expansion of $\Sflow_t$

For any fixed time $t_0$, the operator $\Lop$ may be
decomposed according to
\equation{
  \Lop=\Lopz+\left(t-t_0\right)\Lopp,
  \qquad 
  \Lopp=\sum_{x,\mu}
  \left(\du^a_{x,\mu}S\right)\du^a_{x,\mu}.
  \enum
}
Each term in the power series
\equation{
  \psi_t=\sum_{n=0}^{\infty}\left(t_0-t\right)^n
  \left(\Gopz\Lopp\right)^n\Gopz S
  \enum
}
is then a well-defined differentiable function of both $t$ and $U$.
Moreover, since
\equation{
  \|\Gopz\Lopp\phi\|_{2j}\leq g_{t_0,j}\|\Lopp\phi\|_{2j-2}\leq
  d_jg_{t_0,j}\|\phi\|_{2j-1}\leq d_jg_{t_0,j}\|\phi\|_{2j},
  \enum
}
all $j=1,2,\ldots$ and some constants $d_j$, it is clear that the 
series and all its derivatives with respect to $t$
converge uniformly in the norm $\|\cdot\|_{2j}$
if $t$ is sufficiently close to $t_0$.
Sobolev's lemma (statement (c) in lemma 1.3.5 of 
ref.~[\ref{Gilkey}]) then implies
that the series converges pointwise and uniformly, together
with all its derivatives in $t$ and its derivatives in $U$ up 
to some order proportional to $j$.

If $t$ is in a neighbourhood of $t_0$,
where the convergence of the 
series and its derivatives up to order $l\geq2$
is guaranteed, the action of the 
operator $\Lop$ and the summation may be interchanged
and one finds that
\equation{
  \Lop\psi_t=S-\sum_{n=0}^{\infty}\left(t_0-t\right)^n
  \Popz\left(\Lopp\Gopz\right)^nS
  =\left(1-\Pop\right)S,
  \enum
}
the second equality being implied by the identities $\Pop\Lop\psi_t=0$
and $\Pop\Popz=\Popz$.
As a consequence,
\equation{
  \left(1-\Pop\right)\psi_t=\Gop S=\Sflow_t,
  \enum
} 
which shows that $\Sflow_t$ is,
in the specified neighbourhood of $t_0$,
$l$ times continuously differentiable with respect to $t$ and $U$.
Since $t_0$ and $l$ can be chosen arbitrarily, the 
simultaneous differentiability
of $\Sflow_t$ is thus guaranteed at all times $t$ and to all orders.

\beginbibliography


\bibitem{NicolaiMap}
H. Nicolai,
Phys. Lett. 89B (1980) 341; 
Nucl. Phys. B176 (1980) 419


\bibitem{MoserZehnder}
J. Moser, E. J. Zehnder,
{\it Notes on Dynamical Systems},
Courant Lecture Notes, vol. 12 (American Mathematical Society, 2005)


\bibitem{DelDebbioTauQ}
L. Del Debbio, H. Panagopoulos, E. Vicari,
JHEP 0208 (2002) 044


\bibitem{StefanConference}
S. Schaefer, R. Sommer, F. Virotta,
{\it Investigating the critical slowing down of QCD simulations},
talk given at the XXVII International Symposium on Lattice Field Theory,
Beijing, China (July 2009), arXiv:0910.1465 [hep-lat],
to appear in the proceedings


\bibitem{HMC}
S. Duane, A. D. Kennedy, B. J. Pendleton, D. Roweth,
Phys. Lett. B195 (1987) 216


\bibitem{DuaneEtAl}
S. Duane, R. Kenway, B. J. Pendleton, D. Roweth,
Phys. Lett. B176 (1986) 143;
S. Duane, B. J. Pendleton,
Phys. Lett. B206 (1988) 101


\bibitem{ArnoldI}
V. I. Arnold,
{\it Ordinary Differential Equations}, 3rd ed. (Springer-Verlag, Berlin, 2008)


\bibitem{Gilkey}
P. B. Gilkey,
{\it Invariance Theory, the Heat Equation and the Atiyah-Singer
Index Theorem}, 2nd ed. (CRC Press, Boca Raton, 1995)


\bibitem{Wilson}
K. G. Wilson,
Phys. Rev. D10 (1974) 2445


\bibitem{HairerEtAl}
E. Hairer, C. Lubich, G. Wanner,
{\it Geometric Numerical Integration:
Structure-Preserving Algorithms for Ordinary Differential Equations},
2nd ed. (Springer, Berlin, 2006)


\bibitem{Stout}
C. Morningstar, M. Peardon,
Phys. Rev. D69 (2004) 054501


\bibitem{DDHMC}
M. L\"uscher,
Comput. Phys. Commun. 165 (2005) 199


\bibitem{ReedSimonI}
M. Reed, B. Simon,
{\it Methods of Modern Mathematical Physics}, vol. I 
(Academic Press, New York, 1972)

\endbibliography

\bye